\documentclass[conference]{IEEEtran}
\IEEEoverridecommandlockouts
\usepackage{cite}
\usepackage{amsmath,amssymb,amsfonts}
\usepackage{algorithmic}
\usepackage{graphicx}
\usepackage{textcomp}
\usepackage{mathrsfs}
\usepackage{xcolor}
\usepackage{caption}
\usepackage{subcaption}

\def\BibTeX{{\rm B\kern-.05em{\sc i\kern-.025em b}\kern-.08em
    T\kern-.1667em\lower.7ex\hbox{E}\kern-.125emX}}
\def\endthebibliography{%
  \def\@noitemerr{\@latex@warning{Empty `thebibliography' environment}}%
  \endlist
}
\begin{document}
\title{Safety-Aware Reinforcement Learning for Electric Vehicle Charging Station Management in Distribution Network
\thanks{*Corresponding author: Hao Wang (hao.wang2@monash.edu).}
\thanks{This work was supported in part by the Australian Research Council (ARC) Discovery Early Career Researcher Award (DECRA) under Grant DE230100046.}
}

\author{\IEEEauthorblockN{Jiarong Fan}
\IEEEauthorblockA{\textit{Department of Data Science and AI} \\
\textit{Faculty of IT, Monash University}\\
Melbourne, Australia \\
jiarong.fan@moansh.edu}
\and
\IEEEauthorblockN{Ariel Liebman}
\IEEEauthorblockA{\textit{Department of Data Science and AI} \\
\textit{Faculty of IT, Monash University}\\
Melbourne, Australia \\
ariel.liebman@monash.edu}
\and
\IEEEauthorblockN{Hao Wang*}
\IEEEauthorblockA{\textit{Department of Data Science and AI} \\
\textit{Faculty of IT, Monash University}\\
Melbourne, Australia \\
hao.wang2@monash.edu}

}

\maketitle

\begin{abstract}
The increasing integration of electric vehicles (EVs) into the grid can pose a significant risk to the distribution system operation in the absence of coordination. In response to the need for effective coordination of EVs within the distribution network, this paper presents a safety-aware reinforcement learning (RL) algorithm designed to manage EV charging stations while ensuring the satisfaction of system constraints. Unlike existing methods, our proposed algorithm does not rely on explicit penalties for constraint violations, eliminating the need for penalty coefficient tuning. Furthermore, managing EV charging stations is further complicated by multiple uncertainties, notably the variability in solar energy generation and energy prices. To address this challenge, we develop an off-policy RL algorithm to efficiently utilize data to learn patterns in such uncertain environments. Our algorithm also incorporates a maximum entropy framework to enhance the RL algorithm's exploratory process, preventing convergence to local optimal solutions. Simulation results demonstrate that our algorithm outperforms traditional RL algorithms in managing EV charging in the distribution network.
\end{abstract}


\section{Introduction}
As a clean mode of transportation, electric vehicles (EVs) will play an important role in mitigating climate change. Nevertheless, the widespread adoption of EVs brings substantial challenges to the distribution grid \cite{clement2009impact}. Specifically, uncoordinated charging of EVs can lead to bus voltage drops, thereby impacting the grid reliability. Consequently, the deployment of an intelligent EV charging station management algorithm becomes essential within the distribution network. However, the coordination algorithm faces the challenge of handling multiple uncertainties from renewable energy generation and energy prices. This task demands an effective EV coordination algorithm optimize energy costs while satisfying the voltage constraints and dealing with system uncertainties.

A large body of prior research has focused on developing EV charging management algorithms. By the methodologies, existing studies can be primarily classified into model-based approaches \cite{zheng2018online, beaude2016reducing, luo2019joint} and model-free approaches \cite{duan2019deep,li2019constrained,cao2021smart,tuchnitz2021development,wang2019safe,li2022learning,park2022multi}. The model-based approach relies on the accurate modeling of the system and uncertainties. For example, Zheng et al. \cite{zheng2018online} addressed the uncertainty of EV charging via model predictive control. However, this approach is dependent on the prediction accuracy and could suffer from significant performance degradation with less effective predictions. The non-convex and non-linear characteristics of network models makes the distribution network modeling more challenging. The study in \cite{luo2019joint} relaxed the distribution network model via the second-order cone relaxation technique, but some constraints or dynamic behaviors of distribution networks might not be accurately captured. This approach may oversimplify the model and potentially overlook critical aspects of the network behavior. Model-free strategies, particularly those employing reinforcement learning (RL), have shown significant promise in managing decision-making processes under uncertainties within complex systems. This advancement offers a potential solution for addressing the safe EV charging coordination problem in distribution networks under various uncertainties. For instance, Cao et al. \cite{cao2021smart} developed a customized actor-critic (A2C) algorithm to manage EV charging considering uncertain charging behaviors. In \cite{duan2019deep}, the author used deep RL to learn network features and maintain network integrity by imposing manually designed voltage violation penalties.

Nevertheless, the inherent randomness associated with learning-based algorithms, including RL, brings challenges in satisfying system constraints. To address this problem, \cite{li2019constrained} introduced a safe RL algorithm to solve a constrained Markov decision process (CMDP) problem, thereby avoiding explicit penalties for constraint violations. 
An advanced work \cite{li2022learning} applied the on-policy safe RL in the distribution network, but the on-policy RL requires a large volume of training data and suffers from low data efficiency. Additionally, \cite{wang2019safe} introduced an off-policy safe RL algorithm aimed at reducing the number of voltage violations. However, focusing solely on the frequency of voltage violations may lead to scenarios where fewer but more severe voltage violations occur.

This paper is motivated to bridge the aforementioned research gaps by developing an off-policy safe RL algorithm for coordinating EV charging, which explicitly considers the trade-off between voltage violation number and amount. The studied charging station is integrated with solar energy and can purchase or sell energy through interactions with the grid. The algorithm aims to minimize the total cost while satisfying system constraints under uncertainties. The main contributions of this paper are as follows.
\begin{itemize}
    \item \textit{EV charging management via Safe RL:} 
    In contrast to the conventional method of imposing penalties for constraint violations, our method employs a more systematic approach through the framework of CMDP and using the Soft Actor-Critic-Lagrangian (SACL) algorithm. Our simulation results demonstrate that the Lagrange-augmented safe RL algorithm outperforms conventional RL algorithms in achieving constraint satisfaction.
    \item \textit{Stochastic policy RL with maximum entropy framework:} 
    Our proposed SACL is an RL algorithm that integrates a stochastic policy within a maximum entropy framework, distinguished by its enhanced exploration capabilities. The simulation results demonstrate that our algorithm achieves faster convergence and higher rewards compared to deterministic policy algorithms.
    \item \textit{Trade-off between voltage violation number and amount:} Unlike previous studies only considering the number of voltage violations. Our proposed method considers the explicit trade-off between the number of voltage violation and the magnitude.
\end{itemize}

\section{Problem Formulation}\label{sec:problem}
Our study encompasses multiple EV charging stations, with each station providing services to a set of EVs indexed by $\mathcal{I} {=} {1,..., I }$, each linked to its respective charger within a defined operational timeframe $\mathcal{T}$. Every charging station and its solar energy system are connected to a specific bus in the distribution network. The objective of the system is to minimize the overall energy costs of all charging stations while satisfying network constraints.

\subsubsection{EV Charging Model}
We assume that EV chargers support bidirectional energy flow. The decision variables for each EV are divided into charging power $p_{i,t}^{\text{ch}}$ and discharging power $p_{i,t}^{\text{disch}}$ at time $t$. The constraints associated with these decision variables are as follows
    \begin{align}
        &0 \leq p_{i,t}^\text{ch} \leq (1-z_{i,t})\bar{P}^\text{ch}\label{xt_con},\\
        &0 \leq p_{i,t}^\text{disch} \leq z_{i,t}\bar{P}^\text{disch} \label{xdt_con},\\
        & z_{i,t} \in \{0,1\} \label{z_it},
    \end{align}
where $\bar{P}^\text{ch}$ and $\bar{P}^\text{disch}$ denote the maximum charging and discharging power for EV $i$, respectively. We introduce a binary variable $z_{i,t}$ to prevent charging and discharging actions at the same time for all EVs $i\in\mathcal{N}$ in all times $t\in\mathcal{T}$.

EV charging and discharging will lead to the change in the energy level of EV batteries. The charging station monitors the dynamics of the energy level $E_{i,t}$ of EV $i$ with the associated constraints expressed as follows
\begin{align}    
        &E_{i,t} = E_{i,t-1}+ p_{i,t}^\text{ch} \eta^{\text{ch}}_i \Delta t - \frac{p_{i,t}^\text{disch}\Delta t} {\eta^{\text{disch}}_i} \label{eit},\\
        &A_{i,t} E_i^\text{min}\leq E_{i,t} \leq A_{i,t} E_i^\text{max}, \label{eit_con}
\end{align}
where $\eta^{\text{ch}}_i$ and $\eta^{\text{disch}}_i$ denote the charging and discharging efficiencies for EV $i$, respectively. From the charging station's perspective, the observed EV battery's dynamics can be written in \eqref{eit}. Equation \eqref{eit_con} specifies the minimum and maximum energy levels $E_i^\text{min}$ and $E_i^\text{max}$ for the $i$-th EV's battery with $A_{i,t}$ indicating the connection status of the $i$-th EV at $t$. When $A_{i,t}=0$, EV $i$ is disconnected from the charger and thus the observed $E_{i,t}=0$. To satisfy constraint (\ref{eit_con}), both $p_{i,t}^\text{ch}$ and $p_{i,t}^\text{disch}$ must be set to 0 according to Equation \eqref{eit}, when $A_{i,t}=0$.

When assessing the economic viability of EV charging and discharging, it is vital to consider the additional costs incurred by battery degradation. According to the energy throughput equivalent method \cite{wu2022optimal}, the $i$-th EV's cycle aging denoted $AGE_{i,t}^\text{cyc}$, can be modeled below
\begin{align}
    &AGE_{i,t} = 0.5 \cdot \frac{\lvert p_{i,t}^\text{ch} \eta^{\text{ch}}_i \Delta t -  \frac{p_{i,t}^\text{disch}\Delta t} {\eta^{\text{disch}}_i} \rvert}{E_i^\text{cap}L_i^\text{cyc}}, \label{EFC}
    % &AGE_{i,t}^\text{cyc} = \frac{EFC_{i,t}}{L_i^\text{cyc}}, \label{AGE}
\end{align}
in which $E_i^\text{cap}$ denotes the capacity of the $i$-th EV's battery, while the value of 0.5 represents the average half-cycle of the battery's life. The fraction of battery aging that occurs within a single time step relative to the overall lifecycle of the battery is expressed as $L_i^\text{cyc}$.

\subsubsection{Energy Balance and Network Model}
In our analysis, the distribution network is modeled as a graph $\mathcal{G}$ = \{$\mathcal{N}, \mathcal{E}$\}, comprising a set of nodes $\mathcal{N}_0 = \{0, 1,...,N\}$ and edges $\mathcal{E}$, with node $0$ designated as the substation. The subset of nodes $\mathcal{N} = \mathcal{N}_0 \backslash \{0\}$ represents all nodes excluding the substation node $0$. We denote $n$ is the parent node of $m$. Each node $n \in \mathcal{N}$ is characterized by active power injection $p_n$, reactive power injection $q_n$, and voltage magnitude $v_n$. We adopt the linearized distribution power flow model, specifically the simplified distflow model, which is a linear approximation of the nonlinear branch flow model for radial distribution networks. This approximation is obtained by disregarding the terms for quadratic power loss, and the simplified distflow model is well-established in the area of voltage control. In this model, the variables $p$, $q$, and $v$ satisfy the following equations 
\begin{align}
\begin{split}
   & P_{nm,t} - \sum_{g \in G(n)}p_{ng,t} = \sum_{c \in S(n)} p_{c,t}^{\text{sell},n} \\
   &- \sum_{c \in S(n)} p_{c,t}^{\text{buy},n} + p_{n,t}^\text{V2G} - p_{n,t}^\text{G2V} - p_{n,t}^\text{base}, \label{act}\\
\end{split}
\end{align}
\begin{align}
\begin{split}
   & Q_{nm,t} - \sum_{g \in G(n)}q_{ng,t} = \sum_{c \in S(n)} q_{c,t}^{\text{sell},n} \\
   &- \sum_{c \in S(n)} q_{c,t}^{\text{buy},n} + q_{n,t}^\text{V2G} - q_{n,t}^\text{G2V}- q_{n,t}^\text{base}, \label{react}\\
\end{split} 
\end{align}
\begin{align}
   & v^2_{m,t} = v^2_{n,t} - 2(r_{nm,t}p_{nm,t} + x_{nm,t}q_{nm,t}), \label{volt_cal}\\
   % & |Q^\text{PV}_{n,t}| \leq \sqrt{(S_{n,t}^\text{PV})^2+(P_{n,t}^\text{PV})^2},\\
   % & \sum_{i \in I(n)} p_{ni,t}^\text{G2V} + p_{n,t}^\text{base} = \sum_{c \in S(n)} p_{nc,t}^\text{buy} + p_{n,t}^\text{G2V},\\
   % & \sum_{i \in I(n)} p_{ni,t}^\text{V2G} + p_{n,t}^\text{PVG} = \sum_{c \in S(n)} p_{nc,t}^\text{sell} + p_{n,t}^\text{V2G},\\
   % & \sum_{i \in I(n)} q_{ni,t}^\text{G2V} + q_{n,t}^\text{base} = \sum_{c \in S(n)} q_{nc,t}^\text{buy} + q_{n,t}^\text{G2V},\\
   % & \sum_{i \in I(n)} q_{ni,t}^\text{V2G} + q_{n,t}^\text{PVG} = \sum_{c \in S(n)} q_{nc,t}^\text{sell} + q_{n,t}^\text{V2G},\\
   & \underline{V} \leq v_{n,t} \leq \overline{V} \label{volt_con},
\end{align}
where $G(n)$ represents the set of child buses connected to node $n$ , while $S(n)$ denotes the set of charging stations excluding node $n$. The active/reactive power injection of each node includes energy trading power $p_{c,t}^{\text{sell},n}$/$q_{c,t}^{\text{sell},n}$ and $p_{c,t}^{\text{buy},n}$/$q_{c,t}^{\text{buy},n}$, V2G power $p_{n,t}^\text{V2G}$/$q_{n,t}^\text{V2G}$, grid purchased power $p_{n,t}^\text{G2V}$/$q_{n,t}^\text{G2V}$, and base load $p_{n,t}^\text{base}$/$q_{n,t}^\text{base}$.

Additionally, the voltage level at each bus is assumed to be close to the reference voltage $v_0$, such that \eqref{volt_cal} is rewritten as the following linear form 
\begin{align}
   & v_{m,t} = v_{n,t} - \frac{(r_{nm,t}p_{nm,t} + x_{nm,t}q_{nm,t})}{v_0}. \label{volt_unit}
\end{align}

\subsubsection{Charging Completion}
Each EV must meet its charging requirements before leaving the charging station. The departure time for the $i$-th EV is denoted as $t=dep_i$. Failure to meet the charging requirement will incur a penalty, denoted by $u_i$. For simplicity, \cite{tuchnitz2021development} suggests a linear form for $u_i$ below
\begin{align}
        & u_i = \sigma(E_{i}^{\text{dem}} - E_{i, t}\big|_{t = dep_i} )^+, \label{dem1}
\end{align}
where $\sigma$ signifies a constant coefficient to balance the costs associated with charging and completion of charging tasks. The charging requirement for the $i$-th EV, denoted as $E_{i}^{\text{dem}}$, is characterized as a time-varying parameter.

\subsubsection{Objectives and Constraints}
The operational goals of the charging station service encapsulate two primary objectives: energy cost reduction and satisfaction of charging demands. The objective function, along with operational constraints, is shown as follows
\begin{align}
\begin{split}
&\min ~~ \!\!\!\!\sum_{t\in \mathcal{T}}\Bigl(\sum_{i \in \mathcal{I}} ( AGE_{i,t}^\text{cyc} \kappa^\text{batt} ) \\
&+ \sum_{n \in \mathcal{N}} \bigl((p_{n,t}^\text{G2V} \kappa_t^\text{buy} - (p_{n,t}^\text{V2G} + p_{n,t}^\text{PVG})\kappa_t^\text{sell}) \Delta t\bigr)\Bigr) + \sum_{i \in \mathcal{I}}u_{i}\\
&\text{s.t.}~~\eqref{xt_con} - \eqref{eit_con}~\text{and}~\eqref{act} - \eqref{volt_con},
\end{split}
\label{obj}
\end{align}
where $\kappa^{\text{buy}}_t$ and $\kappa^{\text{sell}}_t$ are energy selling price and energy purchasing price at time $t$. Energy trading revenues and costs offset each other when calculating the overall system cost.

\section{Constrained Markov Decision Process}
We approach the safety-aware EV charging/discharging coordination problem by formulating it as a CMDP and solve it using safe RL techniques. In this context, the distribution system operator (DSO) acts as an agent managing EV charging/discharging together with DERs scheduling within the distribution network, with a crucial focus on maintaining network security. Our CMDP formulation comprises several key components, including state, action, reward, and a auxiliary cost function.

\textbf{State:} The state includes necessary information for the calculation of costs/rewards and constraints. It has two primary segments: EV charging information, including the remaining charging demand $E^r_{i,t}$ and time $T^r_{i,t}$, and the environmental information, including solar power generation $p_t^\text{PVgen}$, electricity selling price  $\kappa_t^\text{sell}$, purchasing price $\kappa_t^\text{buy}$, and bus voltage $v_n$.

\textbf{Action:} The actions in this RL environment include charging/discharging power $p_{i,t}$ for the $i$-th EV at time $t$ and the energy trading decisions $E_{n,t}^\text{trade}$ for the $n$-th charging station at time $t$, which encompass the proportions of the intended trading volumes for each station. Based on these trading actions, the environment will execute the decisions across charging stations, with any remaining power to be sold to or purchased from the grid.

\textbf{Reward:}
The RL algorithm aims to generate an optimal policy by maximizing the model's reward. The objective is to minimize the total system cost while ensuring compliance with system constraints. A negative reward, $R_t$, is employed to capture the total energy cost and user satisfaction, aligning with the objective function in Equation \eqref{obj}. Additionally, an auxiliary cost function is utilized within the CMDP to handle the system constraints.

\textbf{Auxiliary Cost Function: } To avoid manual reward design and adjustment, we tranform the original Markov decision process model to a CMDP model. We specify the auxiliary cost function $c_t$ as follows
\begin{align}
    & c_{t,n} = \sum_{n=0}^{N}[\textbf{1}(|v_{n,t}|>\overline{V})+\textbf{1}(|v_{n,t}|>\underline{V})],\label{volt_num}\\
    & c_{t,a} = \sum_{n=0}^{N}[\text{max}(0,|v_{n,t}|-\overline{V})+\text{max}(0,\underline{V}-|v_{n,t}|)],\label{volt_amount}\\
    & c_t = \beta c_{t,n} + (1-\beta)c_{t,a},
\end{align}
where $c_{t,n}$ represents the number of voltage violations, while $c_{t,a}$ corresponds to the magnitude of these violations. We use $\textbf{1}(\cdot)$ to represent the indicator function. Its value is $1$ when the internal condition is met and 0 otherwise. The upper bound and lower bound of the bus voltage are represented as $\overline{V}$ and $\underline{V}$, respectively. The parameter $\beta \in [0,1]$ is a trade-off weight between the voltage violation number and amount.

\section{Our Proposed Safe RL Method}
In this section, we develop a safe RL algorithm to learn a strategy for managing EV charging stations with the dual objectives of optimizing the cumulative reward and adhering to system constraints within the distribution network. Insufficient exploration of the environment may lead to the agent falling into a local optimum. To avoid this, we employ the Soft Actor-Critic (SAC) method \cite{wang2019safe}, which incorporates a maximum entropy framework to enhance the randomness of action selection and improve the agent's exploration ability. To address system constraints, we use the Lagrange function that integrates network constraints into the value function, eliminating the need for penalty tuning.

\begin{figure}[t]
    \centering
    \includegraphics[width=6.1cm]{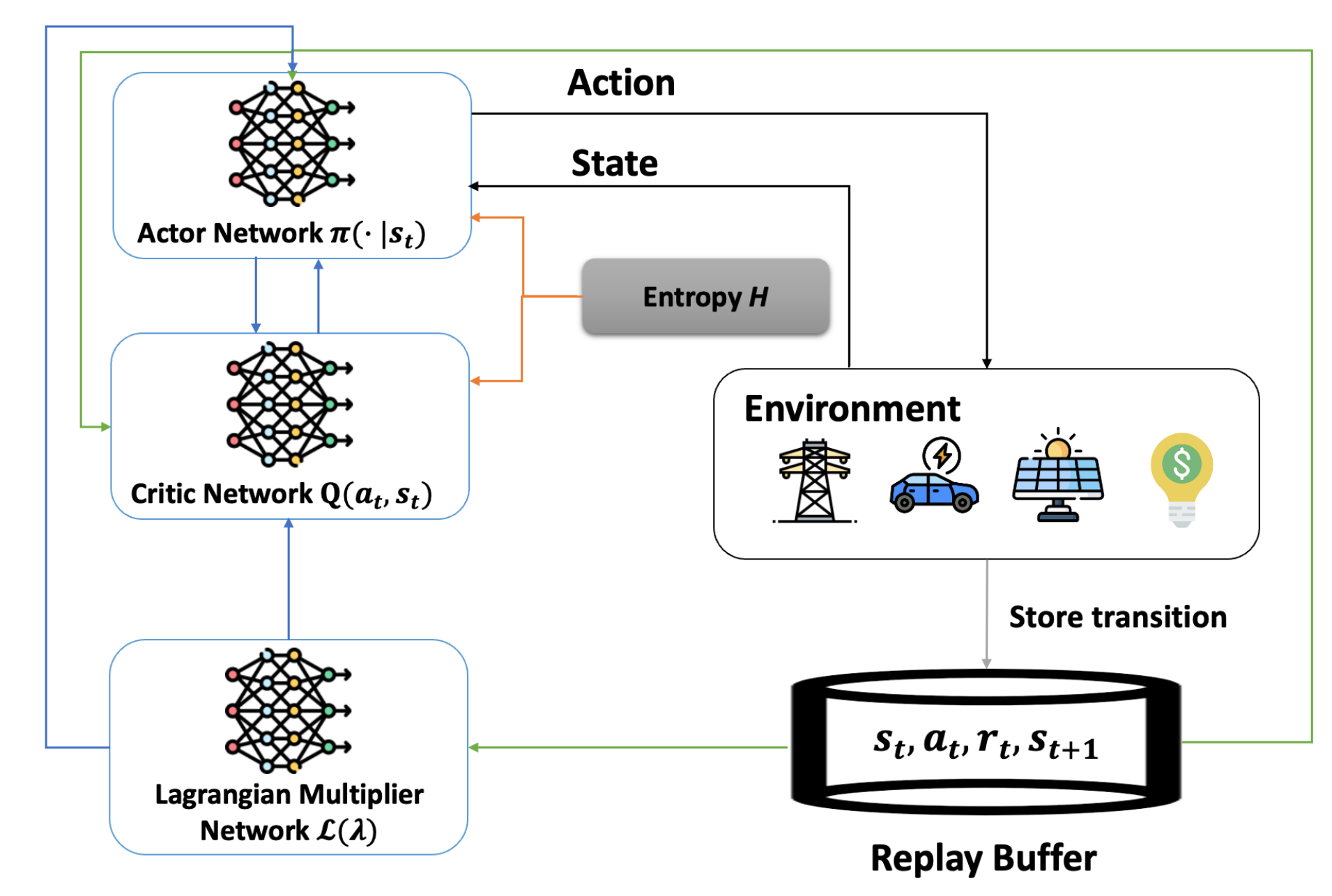}
    \caption{The architecture of the SACL.}
    \label{sacl}
\end{figure}

The SACL framework, depicted in Fig. \ref{sacl}, is based on the A2C architecture. The neural network is used to estimate the function of SACL framework via a data-driven approach. These neural networks share the same architecture with two hidden layers and a rectified linear activation function. Within this framework, the actor network estimates the state-value function and determines the actions for the DSO. Concurrently, the critic network estimates the action-value function and assesses the actions taken by the actor, which aims to reduce the temporal difference error. For Lagrange function, an additional neural network is utilized to process the Lagrange multiplier. The Lagrange network is integrated into our framework to dynamically adjust the penalty associated with constraint violations, ensuring efficient learning without manual tuning of penalty coefficients. This network also features two hidden layers, using ReLU activation functions for hidden layers to support complex function approximation while maintaining computational efficiency.
Within this framework, the goal extends beyond merely maximizing the expected return; it also includes maximizing the entropy of the policy, thereby fostering more exploratory policy behavior. Within the context of maximum-entropy RL, the regularized state-value function is defined as follows
\begin{align}
        & V_\text{soft}(\textbf{s}) = \mathbb{E}_{\tau \sim \pi}[\sum_{t=0}^T\gamma(R_{t} + \alpha H(\pi(\cdot|\textbf{s}_t)))|\textbf{s}_0=s], \label{state_v}
\end{align}
where $H(\pi(\cdot|\textbf{s}_t)))$ represents the entropy of a probabilistic policy at state $\textbf{s}_t$ with $\pi$ serving as the function that maps a state to actions, and $\gamma$ is the discount factor. The action-value function is written as
\begin{align}
\begin{split}
  & Q_\text{soft}(\textbf{s}, \textbf{a}) = \mathbb{E}_{\tau \sim \pi}[\sum_{t=0}^T \gamma R_t \\
  &+ \alpha \sum_{t=0}^T \gamma H(\pi(\cdot|\textbf{s}_t)))|\textbf{s}_0=s, \textbf{a}_0=a].
\end{split}
\end{align}

To deal with system constraints, we define $J_c(\pi)$ as the expected discounted return of the policy $\pi$ with respect to the auxiliary cost function: $J_c(\pi) =  \mathbb{E}_{\tau \sim \pi}[\sum_{t=0}^T \gamma c_t]$. For the SACL framework, the optimal policy of CMDP can be obtained by solving
\begin{align}
    & \max_\pi \mathbb{E}_{\textbf{s} \sim \textbf{D}}[V_\text{soft}(\textbf{s})], \; s.t \;J_c(\pi) < d, \label{CMDP}
\end{align}
where $\textbf{D}$ is the historical data replay buffer, and $d$ is a small tolerance for the voltage constraint violation. In order to solve \eqref{CMDP}, the optimization problem with constraints can be converted to a Lagrange function written as
\begin{align}
    & \mathcal{L}(\pi,\lambda) = \mathbb{E}_{\textbf{s} \sim \textbf{D}}[V_\text{soft}(\textbf{s})] + \lambda(d - J_c(\pi)),
\end{align}
where $\lambda$ is the Lagrange multiplier to be updated at each iteration. The updating rules for value functions and $\lambda$ are presented in \cite{wang2019safe}.

\section{PERFORMANCE EVALUATION} \label{sec:eval}
We take the perspective of the DSO to train an agent for managing a distribution network and use the IEEE 33-bus system in the simulations. We use PandaPower to build the simulation system using its default base load data. Four charging stations are placed at node 8, node 12, node 22, and node 30. Each charging station has five EV chargers with a charging power of 22kW and solar panels with a capacity of 13kWp. EV charging data are obtained from \cite{lee2019acn}. EVs purchase energy following Time-of-Use (TOU) pricing, while energy is sold to the grid at wholesale energy prices \cite{aemo_dashboard}. The energy trading price is set as the midpoint of the TOU energy price and the wholesale market energy price. Solar energy, which is free to use, is simulated using solar energy generation data from \cite{Elia_2021}.

To effectively manage charging stations and meet system constraints, we implement the SACL algorithm, handling system constraints through the application of the Lagrange function. SACL enhances environmental exploration and prevents convergence on suboptimal local solutions via its stochastic policy embedded within a maximum entropy framework. The hyper-parameters of SACL are determined through the empirical testing to identify configurations that yield the best performance across various simulation scenarios. For comparative analysis, we also consider Deep Deterministic Policy Gradient (DDPG) and SAC algorithms, as DDPG is a representative deterministic policy algorithm and SAC employs penalties for constraint violations.

\begin{figure}[t]
    \centering
    \includegraphics[scale=0.15]{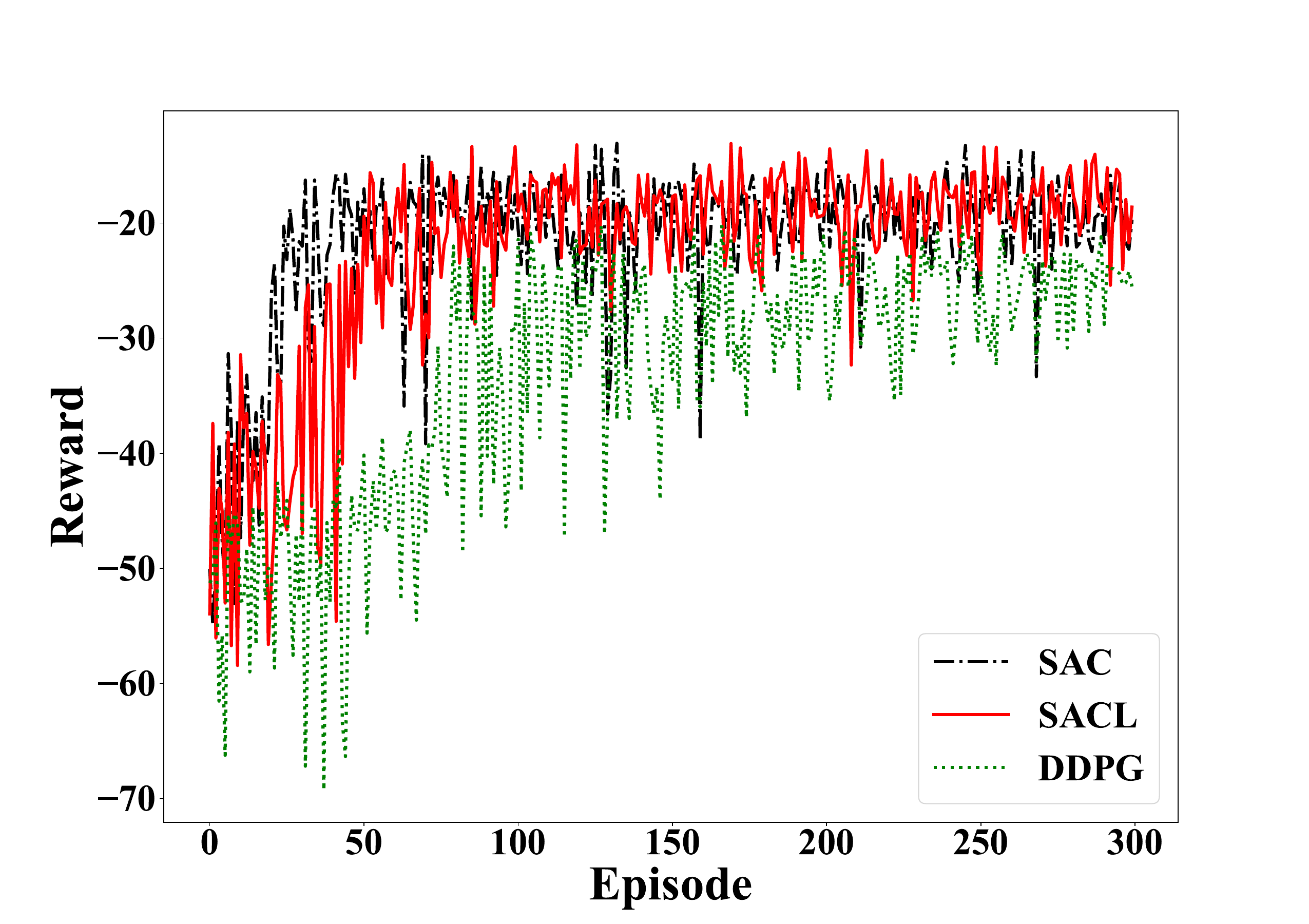}
\caption{The reward convergence for different algorithms.}
\label{reward}
\end{figure}

Fig. \ref{reward} illustrates the convergence of rewards for these algorithms. All three algorithms demonstrate convergence within 100 episodes. SAC and SACL exhibit faster convergence and achieve higher rewards compared to DDPG, attributed to the maximum entropy framework that expedites the exploration of more effective policies.

\begin{figure}[t]
    \centering
    \includegraphics[width=8.8cm]{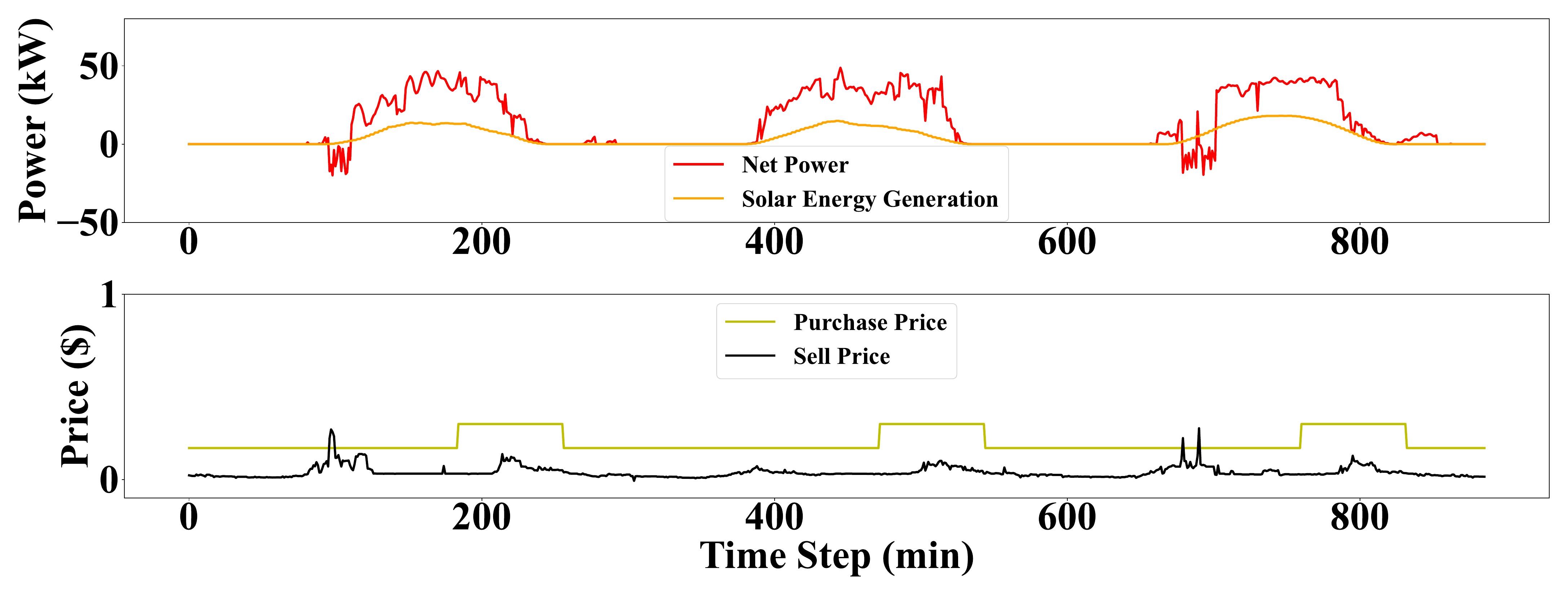}
    \caption{Net power of one charging station under uncertainties of solar energy and real-time price.}
    \label{compare}
\end{figure}

Fig. \ref{compare} shows the net power output at a single charging station under various uncertainties. It is clear that our proposed algorithm successfully captures the pattern of solar energy generation, optimizing the use of freely available solar energy. Furthermore, the algorithm guides the charging station to sell energy back to the grid when the selling price exceeds the purchasing price. The results depicted in Fig. \ref{compare} demonstrate the effectiveness of the proposed algorithm.

\begin{figure}[t]
    \centering
    \includegraphics[scale=0.23]{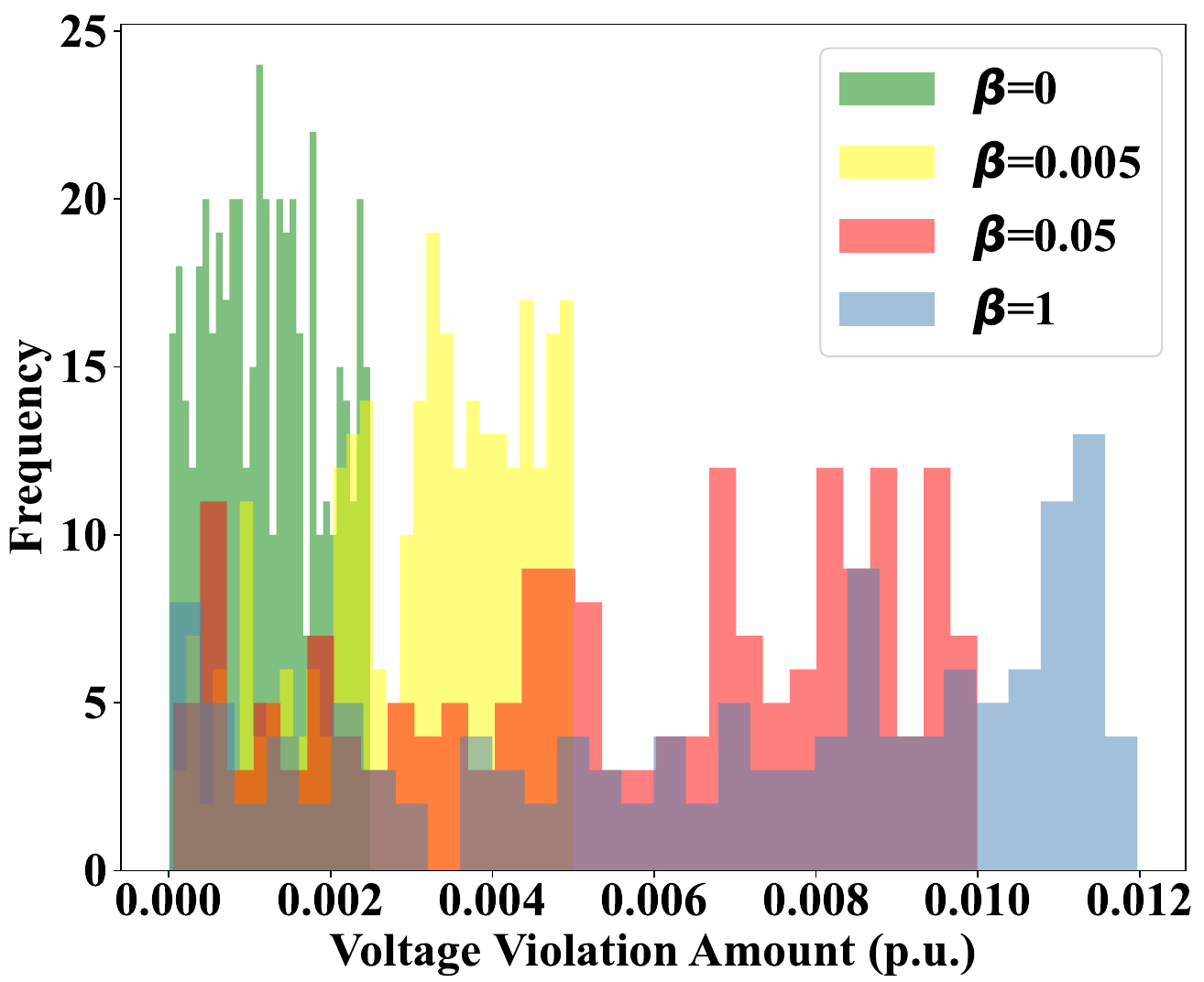}
\caption{The frequency histogram for voltage violation amount.}
\label{volt_fre}
\vspace{-3mm}
\end{figure}

To regulate bus voltage, we consider two parameters: voltage violation number (VVN), as defined in \eqref{volt_num}, and voltage violation amount (VVA), outlined in \eqref{volt_amount}. Focusing solely on VVN can lead to a few severe voltage violations, while an exclusive emphasis on VVA could result in numerous minor violations. Therefore, achieving a balance between VVN and VVA is crucial. Fig. \ref{volt_fre} illustrates this trade-off through a frequency histogram. As the trade-off parameter $\beta$ decreases, the emphasis shifts towards VVA, resulting in an increased number of violations. Conversely, a higher value of $\beta$ ($\beta=1$) reduces VVN but may lead to more severe voltage violations.

Table \ref{pe} compares energy cost reduction and voltage satisfaction performance among different algorithms. With consistent parameter settings, SAC and SACL outperform DDPG in energy cost reduction and charging demand satisfaction, due to their superior exploration capabilities. Regarding voltage constraints, the VVN and VVA of safety-aware SACL are lower than those of SAC, underlining SACL's effectiveness in satisfying voltage constraints.
\begin{table}[ht]
\centering
\caption{Performance comparisons between SACL, SAC, and DDPG in terms of cost, unfinished demand (UD), voltage violation number (VVN), and voltage violation amount (VVA).}
\begin{tabular}{||l|l|l|l|l||}
 \hline
Algorithm     & Cost (\$) & UD (kWh) & VVN & VVA (p.u.)  \\
DDPG          & 1091      & 412.7     & 551 & 1.611       \\
SAC           & 933       & 427.5     & 436 & 1.284       \\
\textbf{SACL} & 926       & 411.2     & 294 & 0.915      \\
 \hline
\end{tabular}
\label{pe}
\end{table}

\section{Conclusion and Future Work}
Our work developed an innovative safety-aware reinforcement learning approach for managing multiple EV charging stations in distribution networks. Our proposed algorithm excels in striking a balance between cost optimization and adherence to system constraints, particularly under uncertainties including fluctuating energy prices and solar energy generation. Our simulation results demonstrate that our algorithm outperforms widely used reinforcement learning methods by leveraging the maximum-entropy framework and the Lagrangian function. Future research will consider reactive power for enhanced voltage control and seek to improve the scalability of the algorithm to coordinate more charging stations in larger distribution networks.

% References
\bibliographystyle{IEEEtran}
\bibliography{IEEEabrv.bib,ref.bib}

\end{document}